\documentclass[11pt,twoside]{article}

\usepackage{asp2006}
\usepackage{epsf}

\begin{document}
\title{The New Science of Gravitational Waves}   
\author{Craig J. Hogan}   
\affil{University of Washington, Seattle}    

\begin{abstract} 
A brief survey  is presented of   new  science that will emerge during the decades ahead from direct detection of  gravitational radiation.  Interferometers on earth and in space will probe the universe in an entirely new way by  directly sensing motions of distant matter over a range of more than a million in frequency. The most powerful sources of gravitational (or indeed any form of)  energy in the universe are  inspiralling and merging binary black holes;  with   LISA data, they will become the most distant, most completely and precisely modeled, and most accurately measured systems in astronomy outside the solar system. Other sources range from already known and named nearby Galactic binary stars,  to compact objects being swallowed by massive black holes, to possible effects of new physics: phase transitions and superstrings from the early universe, or holographic noise from quantum fluctuations of local spacetime.\footnote{Parts of this survey are based on  text prepared by the author for the executive  summary of LISA science in the LISA mission's report to the NRC's Beyond Einstein Program Assessment Committee \citep{LISAscience}, where  more extensive references can be found. Other reviews and summaries can be found in \citet{flan}, \citet{amsci}, \citet{hughes03}, \citet{hughes06},  \citet{LISA6}, \citet{thorne},  and \citet{schutz}. }
\end{abstract}



\section{The Sounds of Spacetime}
The grand historical perpective of this 50th anniversary of NRAO recalls many examples of how new kinds of data about the world drive transformations in science.  Radio astronomy revealed the  largely nonthermal emissions of low frequency light, unveiling new aspects of the universe that were previously not even guessed at.  The range of radio phenomena now studied is enormous, from the vicinities of neutron stars and black holes, to the birthplaces of stars, to the magnetized plasmas and relativistic particles in intergalactic space--- a range of over twenty orders of magnitude in linear scale. Radio astronomy has given us a vastly broader and  deeper appreciation  of the  activity of the cosmos, and
 promises more great advances in the decades to come.

 In this survey I  anticipate   another  dramatic impending transformation in  knowledge, from  the direct detection of gravitational waves.  This new science will bring a qualitatively  new way of probing the distant universe--- like adding a sound track for the first time to the electromagnetic picture of the universe which, however richly textured and vast in scope,  has up to now been  silent.

Einstein's theory of spacetime and gravity, general relativity, predicts that motions of mass produce propagating vibrations that travel through spacetime at the speed of light. These gravitational waves (as the vibrations are called) are produced abundantly in the Universe and permeate all of space; they penetrate all matter, back to the Big Bang and down to the event horizons of black holes; but they have never been detected directly.\footnote{Nevertheless, the existence of gravitational waves is in little doubt as their effects  have been measured precisely, if  indirectly.  The long standing best evidence \citep{binary} for gravitational waves is the orbital decay of  the Hulse-Taylor binary pulsar, which radiates at frequencies close to  the operating band of planned detectors such as LISA.}
  Measuring gravitational waves directly  will add an altogether new way to explore what is happening in the Universe: rather than studying the propagation and transformation of conventional particles and fields in spacetime, as all science has done up to now, we will directly  sense vibrations of the fabric of spacetime itself produced by the motion of distant matter.  Studying this new form of energy will convey rich new information about the behavior, structure,  and history of the physical universe, and about physics itself.

\section{Detectors and Sources}

This transformational new science is happening due to the technical capabilities of detection, in the same way that   telescopes and   microscopes in their time opened new kinds of phenomena to study.  Gravitational wave detectors \citep{LIGO} are supersensitive microphones that measure motions caused by spacetime vibrations. The overall gravitational wave energy budget of the universe (in terms of power or energy density) is comparable with the electromagnetic energy budget, but spacetime is an extremely stiff medium so even the large amounts of energy expected in cosmic gravitational waves  produce only tiny distortions.  Even so, during the next decade detectors are expected to attain the level of sensitivity required to detect many kinds of distant motions. 

  The frequencies of the waves correspond to the physical motions that produce them.    Capabilities are being developed to detect waves from kilohertz down to below a millihertz using direct interferometry, and down to  a  nanohertz using millisecond pulsar timing. As in electromagnetic astronomy, that enormous range in frequencies maps onto a large range of astrophysical systems.

Laser interferometer detectors measure the relative motions of proof masses. From the ground, detectors such as LIGO, VIRGO, GEO and TAMA probe frequencies of about 30 to 1000 Hz.  The strongest sources of waves in that band are expected  to be the final stages of coalescences and mergers of stellar-mass binary black  holes.  In the next decade, these detectors will reach far enough into space to expect an appreciable rate of these rather rare events.

LISA is a proposed space mission designed to measure gravitational radiation over a broad band at low frequencies, from about 0.1 to 100 millihertz, a band where the Universe is richly populated in a wide variety of strong sources of gravitational waves. This rich activity at low frequencies  makes sense from an intuitive point of view since after all, astronomical systems are physically large and even with very high velocities most of them do not change very  quickly.  (In addition, waves from sources at high redshift are slowed further by cosmological time dilation).  Indeed, at the low end of its band LISA will be limited by a confusion-limited background of gravitational waves from compact binary stars \citep{farm}.  (LISA will  detect the progenitor populations of the binaries that eventually and rarely flare up as LIGO sources.)

Low frequency signals come from a wide range of different sources:  massive black holes merging in galaxies at all distances; massive black holes consuming smaller compact objects; known binary compact stars and stellar remnants; members of known populations of more distant binaries; and probably other sources, possibly including relics of the extremely early big bang, that are as yet unknown.  These strong signals convey detailed information addressing a wide range of science:  the history of galaxies and black holes in the Universe; general relativity itself and the behavior of spacetime; precision measurements of the Universe as a whole; the physics of dense matter, stellar remnants and compact binaries; and possibly new physics associated with events in the early Universe,  relics predicted in string theory, or even direct detection of quantum gravitational noise in the detector.

In the same way that electromagnetic radiation accompanies acceleration of electric charges, gravitational radiation accompanies quadrupolar acceleration of any kind of mass or energy.  Strongly quadrupolar motion in a system with mass $M$  and size  $R$  at a distance  $D$ typically perturbs spacetime with a dimensionless metric-strain amplitude (fractional variation in proper spatial separations) of about  $h\simeq (GM/Rc^2)^2 (R/D)$.  Interferometric detectors sense  this distortion by monitoring the changes in the distances between inertial proof masses.   LISA in particular will use precision laser interferometry across a vast distance of space to compare separations among proof masses that are protected by the spacecraft from non-gravitational disturbances.  Over a separation of $L\approx 5$  million kilometers, LISA's ability to measure variations of $\Delta L\simeq 0.05$ picometers corresponds to a strain amplitude sensitivity of about  $h\simeq \Delta L/L\approx 10^{-23}$. LISA coherently measures spacetime strain variations, including frequency, phase, and polarization, all of which reflect large-scale properties of the systems that produce them and are therefore direct traces of the motions of distant matter.

These detectors have an all-sky field of view, but  coherent modes of observing  resolve and distinguish overlapping signals and locate them on the sky.  Large dynamic range  allows study of sources within the Galaxy and out to the edge of the Universe.   The  wide frequency band (more than three decades for LISA alone) allows them to study similar sources of widely different masses and cosmological redshifts. Because gravitational waves penetrate all regions of time and space with almost no attenuation,   waves reach us from the densest regions of matter, the earliest stages of the Big Bang, and the most extreme warpings of spacetime near black holes.

\section{Survey of LISA Science}

Although the first gravitational wave detection will likely come   from the ground-based interferometers (or possibly from millisecond pulsar timing), in the course of the next twenty years we can hope to build an interferometer in space, such as LISA,  which will open up the rich low frequency gravitational wave domain.  This is a brief summary of the  range of opportunities in LISA's low frequency band,  0.1 to 100 millihertz, and sensitivity, $h\approx 10^{-23}$, as viewed from 
the electromagnetically limited perspective of 2007.

\subsection{Massive Black Hole Inspiral, Coalescence, and Merger}

LISA will record the inspirals and mergers of binary black holes, the most powerful transformations of energy in the Universe, allowing precision measurements of systems composed only of pure dynamical spacetime.

The strongest gravitational waves are generated by systems with the largest gravitational potential $GM/R$, hence large masses and small sizes.  The strongest of all are generated by interactions of black holes, dense knots of pure spacetime energy with $GM/Rc^2\approx 1$.  At LISA frequencies the strongest sources are massive black hole binaries with about $10^4$  to $10^7$  times the mass of the sun.  Two black holes orbit each other, spiral together as they lose energy by radiation, and finally merge. The waves from these events--- many cycles over a long inspiral, climaxing in a brief series of powerful waves during a violent merger, and a final ringdown to a quiescent single black hole--- record dynamical general relativity in its purest form but also in its most violent, nonlinear dynamical behavior: a maximally warped vacuum spacetime interacting with itself. 

The black hole binaries start with wide orbits at low frequencies. As they lose energy their frequency increases and their radiation strengthens.  A typical source enters the LISA band a year or more before the final merger so many orbits are recorded, encoding details of the system properties and behavior, position on the sky, and absolute distance.  The coherent phase and polarization information obtained over LISA's solar-orbit baseline (and variable inclination) can often pinpoint where a source is in the sky to better than a degree.  In the last hours or minutes the signal-to-noise ratio grows very high, often into the hundreds to thousands depending on distance.  At its peak luminosity, around the moment of merger, a black hole binary (of any mass) is the most extreme transformation of mass-energy of any kind in the Universe, radiating a power  of about $10^{-3}c^5/G$ (or $10^{49}$  watts), in a few wave cycles, spread over a time of about 100 Schwarzchild times $GM/c^3$.  The  peak radiated power from any comparable-mass black hole merger, from just one source, is about 1000 times more than all the starlight power  in the visible Universe.  The merger throes of a million solar mass binary black hole merger last about 500 seconds.  

Massive black hole binary inspiral and merger events are such powerful radiators that LISA can detect them anywhere, out to the largest redshifts where galaxies might exist. The LISA signals during the merger phase are so strong that the signal-to-noise ratio is often greater than 100 even in one oscillation cycle: signal waveforms are visible on an oscilloscope type display of raw data even to the naked eye, so even if general relativity were to be wrong at the levels allowed by our existing tests (e.g. the binary pulsar) we would be able to use LISA data to make sense of what is happening.

The detailed study of waveforms from black hole binaries  tests the nonlinear behavior of spacetime predicted by general relativity at high precision.  Recent breakthroughs \citep{pretorius,baker1,Campanelli:2005dd,cent} now allow numerical computation of Einstein's field equations throughout the entire inspiral and merger event, yielding a detailed map of the predicted gravitational waveform that will be the first detailed test of dynamical, strong-field general relativity.  Waveforms coherently correlated over many orbits (from $\approx 10$ to $\approx 1000$  depending on mass and redshift) recorded in the LISA signal stream, and detection of events with a signal to noise of a thousand or more, allow   precise tests of the theory as well as precise measurements of all system parameters to a precision of order $10^{-2}$   to $10^{-3}$ from gravitational physics alone. Comparison with the computed details of the inspiral and merger waveform verifies the agreement with the binary black hole model of the system, including details of the self-interaction of spacetime.

\subsection{Precision Relativity and Extreme Mass Ratio Inspirals}

LISA will map isolated black holes with high precision, verifying that they are the stationary Òno hairÓ spacetime configurations described by the Kerr metric, completely specified by four numbers: the mass and three components of spin.

In general relativity an isolated spinning black hole is described mathematically as a particular, precisely specified  spacetime shape called a Kerr metric, that depends only on the physics of gravity and not at all on the history or environment of the black hole.  Comparison of the ringdown waveform after a merger event with theory verifies that the final system is indeed described by the Kerr solution, and satisfies the Òno hairÓ theorem of general relativity which states that an isolated, stationary black hole is completely specified by its mass and angular momentum.  

 LISA also uses  extreme mass ratio inspirals (EMRIs) to explore the spacetime near a massive black hole. Driven by chance encounters, a much smaller mass compact object--- such as a degenerate dwarf, neutron star or stellar-mass black hole--- sometimes finds itself captured by the massive black hole, after which it orbits  many times until it finally plunges into the horizon and disappears. Based on extrapolation from conditions near the center of our Galaxy, it is estimated that hundreds to thousands of these events may be detected by LISA.
 The gravitational waves from these sources encode a detailed map of a relatively unperturbed massive black hole, predicted to be a pure Kerr knot of highly curved, spinning spacetime.  About  $10^5$ wave cycles are measured for each source, emitted from orbital paths exploring deep into different parts of the relativistic region near the massive black hole.  The specific mass quadrupole and higher moments predicted by the Kerr solution are measured with a precision of about $10^{-4}$, and precision tests of small variations about the equilibrium Kerr solution---  the small amounts of ``hair'' added by the perturbing object--- are measured at the one percent level \citep{bc1,bc2}. Gravitational waves from these events map in exquisite detail the cleanest and most accurately predicted structures in all of astrophysics, whose mathematical elegance Chandrasekhar once likened to that of atoms.

\subsection{History of massive black holes and galaxy formation}

LISA will directly observe how massive black holes form, grow, and interact over the entire history of galaxy formation \citep{vol}.

Optical, radio and x-ray astronomy have produced abundant evidence that nearly all galaxies have massive black holes in their central nuclei (and indeed that some recently merged galaxies even have two black holes).  These nuclear black holes have a profound effect on galaxy formation; the influence of black hole powered jets on the intergalactic gas out of which galaxies form is in some cases directly observed.  There is a circumstantial case, but no direct evidence, that the formation of this population of black holes was associated with a multistage process of binary inspiral and merger, together with accretion.  LISA will obtain direct and conclusive evidence and study details of this process via gravitational radiation.

In standard concordance cosmology, the first massive black holes naturally arise from the very first, supermassive stars.  In this scenario, black hole binaries begin to form from a high redshift,  $z\approx 20$,  when galaxies start to assemble by a series of (hundreds to thousands of) hierarchical mergers of smaller protogalaxies.  When two galaxies merge into one, their central black holes sink to the center of the new galaxy, find each other, inspiral and merge.  There are so many galaxies forming in the Universe observed by LISA that mergers happen quite frequently: estimates based on standard galaxy formation theory suggest that if black holes indeed grew by hierarchical merging, LISA detects a merger event about once or twice every week on average, from a wide range of redshifts extending back to massive binaries in early protogalaxies at $z\approx 15$. At any given time, in addition to the actual mergers these models predict that LISA observes inspiral signals from hundreds of binaries in the final years before their merger. LISA digs directly and intimately into the detailed evolution of galactic nuclei: the large sample of binaries provides a direct record of the whole history of galaxy formation in the observable Universe, and of the processes that grew their central black holes and shaped their nuclei.\footnote{The predictions of massive nuclear black hole merger event rates from hierarchical galaxy formation are uncertain since no direct data exist as yet to constrain the detailed assumptions of the models of evolution of galactic nuclei. Other models, where massive black holes grow mainly by accretion, yield much lower rates, though still observable;  either way, LISA will uncover mechanisms of massive black hole formation. Similarly, rates estimated for extreme mass ratio inspiral events depend on extrapolation from a  small sample of nearby galaxies where nuclear stellar populations can be studied directly,  yielding estimates on the order of 100 to several hundred events per year.}   In addition, the parameters measured from extreme mass ratio events yield a census of isolated massive black hole spins and masses in many galaxies today, a revealing relic of black hole history.   The local universe also produces observable  inspirals of  less compact stars and stellar remnants that  probe the rich astrophysics of  massive central black holes consuming piecemeal the various stellar populations in their vicinity.  

\subsection{Precision Distances and Cosmology}

LISA will measure precise, gravitationally-calibrated absolute luminosity distances, with the potential of contributing uniquely to measurement of the Hubble constant, cosmological geometry, dark energy, and neutrino masses.

Because the inspiral leading up to a black hole merger is a clean, pure vacuum spacetime system, properties of the radiation can be computed exactly in general relativity, so that the black hole masses, spins, orientations and even the exact distance and direction can be reconstructed from LISA data.  (Roughly speaking, the final wave cycle period tells the final absolute Schwarzschild radius, and the ratio of that length to the distance is the metric strain, $h$.)  These inspiral distances are both individually precise and absolutely calibrated, using only pure gravitational physics, and they cover a wide range of redshift.  In the absence of propagation effects the absolute physical luminosity distance to a single LISA inspiral event is typically estimated from the waveform alone with about one percent, and in some cases with as good as 0.1\% precision.  If identification of the host galaxy,\footnote{LISAÕs waveform fitting can often pinpoint the direction of a source to much better than a degree, and the distance estimate also narrows the redshift range considerably; nevertheless at high redshift there may be many thousands of galaxies in the LISA Òerror boxÓ  for a given source.  Models suggest that the host may be identified from a telltale nuclear starburst associated with the merger, or from variability associated with the disrupted disks around the merging holes,(e.g., \citet{milo}) although galaxy nuclei are too little understood to make  a firm prediction.  It is possible that identification of hosts will prove elusive. } or a statistical estimate of redshift from a sample of galaxies in the allowed range of LISA-determined positions,\footnote{Since galaxies cluster strongly, redshift surveys can provide significant redshift information even in situations where the specific host galaxy cannot be identified.}   allows an independent redshift determination, the redshift-distance relation is also measured with high precision \citep{schutz86}. 
  Black hole binaries thus represent a unique and independent new capability for precision cosmology that complements other techniques. 

 Even a small number of sources at moderate redshift calibrates the distance scale and Hubble constant  better than any current method--- a powerful and complementary constraint on cosmological  models in combination  with microwave background data, baryon acoustic oscillations, supernovae, weak lensing or other techniques.  The expected   sample   inspiral events may lead to measurements of cosmological parameters comparable in precision to other methods, but with independent and absolute (and purely gravitational) calibration,   and completely different systematic errors. The main source of error at high redshift  is the noise induced by cosmic weak gravitational lensing along the line of sight. To some extent in  a statistical sample this is controllable, and indeed provides unique new information about the nature and clustering of dark matter over time.
 
\subsection{Binary Compact Stars}

LISA will study in detail thousands of compact binary stars in the Galaxy, providing a new window into matter at the extreme endpoints of stellar evolution. 

In addition to mergers and meals of distant black holes, LISA detects many lower mass binary systems in our Galaxy, mostly   compact degenerate remnants of normal stars.  Very soon after turning on, LISA will quickly detect a handful of nearby binary white dwarf stars already studied and named.  These Òverification binariesÓ provide sources with known positions and periods, ensuring particular, predictable LISA signals.  Signals are also certain to appear from populations in our galaxy of numerous and various remnants, including white dwarfs and neutron stars, which are known to exist from small samples known  to emit electromagnetically.  Simple extrapolation of known nearby samples to the whole Galaxy predicts that LISA will detect  and fit in detail  thousands of binaries. The most compact binaries (those at high frequency) will be measured in detail as individual sources from across the Galaxy, while at lower frequencies only the nearby ones will be individually distinguished; millions of others from across the Galaxy will blend together into a confusion background that limits LISA's low frequency sensitivity.  LISA provides distances and detailed orbital and mass parameters for hundreds of the most compact binaries, a rich trove of information for detailed mapping and reconstruction of the history of stars in our galaxy, and a source of information about tidal and other nongravitational influences on orbits associated with the internal physics of the compact remnants themselves.  LISA may also detect at high frequencies the background signal from compact binaries in all the other galaxies \citep{farm}.

\subsection{New Physics: Phase Transitions and Cosmic Superstrings}

LISA may find entirely new phenomena of nature not detected using light or other particles \citep{newphysics}.

Given that all forms of mass and energy couple to gravity, other sources of gravitational waves may exist that are not known from extrapolating current electromagnetic observations.  LISA's frequency band can even be extrapolated to very high redshift where we do not yet have any direct observations, and to a regime where LISA itself will be our first information of any kind about the nonlinear behavior and motion of matter.  For example, the LISA frequency band in the relativistic early Universe corresponds to horizon scales at the Terascale frontier currently being explored at CERN and Fermilab. Whereas accelerators study single-particle collisions,  LISA will explore collective behavior of fields that often accompany symmetry breaking--- phase transitions of new forces of nature or extra dimensions of space that may have caused catastrophic, explosive bubble growth and efficient gravitational wave production. LISA is capable of detecting a stochastic background from such events with critical temperatures  from about 100 GeV to about 1000 TeV, if gravitational waves in the LISA band were produced with an overall efficiency more than about $10^{-7}$, a typical estimate from a moderately strong relativistic first-order phase transition.  This corresponds to cosmic times about $3\times 10^{-18}$  to $3\times 10^{-10}$ seconds after the start of the Big Bang, a period not directly accessible with any other technique. 

Reaching much farther still beyond the range of any particle accelerator, LISA also deeply probes possible new forms of energy such as cosmic superstrings \citep{pol},  relics of the early Universe predicted in some versions of string theory that are invisible in all ways except by the gravitational waves they emit.  In principle, their signature could provide direct evidence for new ideas unifying all forms of mass and energy, and possibly even spacetime itself.\footnote{It should be noted that currently, the best observational constraints on cosmic superstrings already come from gravitational wave data--- not from an interferometer, but from the lack of noise in the residuals in times of pulse arrivals from millisecond pulsars \citep{DePies:2007bm}.  Future development of millisecond pulsar techniques--- better timing, and bigger samples, such as the Pulsar Timing Array--- promise  significant improvement. It is guessed that the current data are within an order of magnitude of detecting gravitational waves from massive black hole binaries (long before merger), so this technique has a reasonable possibility of delivering the first direct detection of gravitational waves.}

\section{Holographic Noise}

In addition to the classical gravitational waves predicted by standard general relativity, it is possible \citep{Hogan:2007rz,Hogan:2007hc} that spacetime on its own produces quantum noise detectable in laser interferometers.  This is not  predicted by classical relativity but is a generic effect in    ``emergent'' spacetimes, where  spacetime is not fundamentally a smooth manifold, as assumed in general relativity and quantum field theory, but  only emerges as a classical limit of a more fundamental theory based on quantum wave mechanics.  If the particle paths that define a spacetime metric are themselves defined as observables of quantum waves of Planck wavelength $l_P$, there is an unavoidable fuzziness to spacetime.  Locally blurring occurs only at the Planck scale, but in an extended region of size $L$ it is much larger, of order $\Delta x \approx \sqrt{l_PL}$. It can be thought of as a quantum  indeterminacy of angles,  of magnitude $h\approx \sqrt{l_P/L}$,  caused by the diffraction limit of Planck scale waves. 

One reason to believe that quantum gravity might behave this way is that it correctly predicts  ``holographic'' behavior.  It has been previously noticed in black hole physics and string theory that the number of degrees of freedom of a system in spacetime only increases with area and not with volume: like a hologram, the behavior of any whole system can be encoded on a 2D surface.  Viewed ``from inside'', a spacetime with such a limited information content should display holographic indeterminacy.

The angular distortions are small even by the standards of radio astronomy, but are within reach of  interferometers. 
Holographic indeterminacy leads to a quantum noise   with a shear-like spatial behavior and a universal spectrum of metric perturbations,
$h_{rms} \simeq \sqrt{l_P/c}\simeq 2.3 \times 10^{-22} /\sqrt{\rm Hz}$.  This ``holographic noise'' may be detectable with certain interferometer designs, including gravitational wave detectors such as LISA that measure distances around a triangle.  LIGO's geometry   does not see it since the holographic noise is entirely transverse to separation; on the other hand, the sensitivity of LIGO would easily detect the effect if it were deployed in a different (LISA-like) layout.   
If holographic indeterminacy is found  it will allow the first direct experimental study of quantum gravity.

\section{Gravitational Waves and Electromagnetic Astronomy}

Study of   electromagnetic counterparts to gravitational wave sources will provide an exploratory bonanza for wide field imaging and spectroscopy across the electromagnetic spectrum, especially with the addition of synoptic or time-domain surveys.   Many of the new Galactic binaries discovered with LISA will be observable electromagnetically.  Precursors and afterglows of accretion disks associated with black hole mergers, or burps associated with black holes dining on stellar populations, typically  produce varying thermal sources in the rest-frame UV, but a wide variety of nonthermal and processed radiation will appear from radio to gamma rays.  
Telescopes such as  the proposed Large Synoptic Survey Telescope will make  movies of LISA black hole merger events   with a field of view encompassing LISA's error boxes and  time sampling comparable to  LISA's,  monitoring synchronously the variations in emission from accretion disks during the final stages of binary black hole inspiral and merger,   at the same time that LISA maps in detail the changes in the local gravitational field using gravitational waves. 
Gravitational wave astronomy will become part of the routine repertoire of coordinated observations that dig deeply  into the precise physical characteristics of complex and extreme astrophysical systems.


\acknowledgements 
The author is grateful for support from the Alexander von Humboldt Foundation and the Max-Planck-Institut f\"ur Astrophysik.


\end{document}